\documentstyle[preprint,aps,psfig]{revtex}
\begin{document}
\draft
\preprint{}
\title{Quantum Vacuum in Hot Nuclear Matter -- A Nonperturbative Treatment}
\author{Amruta Mishra$^{\dagger}$, P.K. Panda\footnote[1]
{email: panda@ift.unesp.br}$^\ddagger$, 
W. Greiner$^\S$}
\address{$^\dagger$ Institute for Plasma Research,
Gandhinagar - 382428, India} 
\address{$^\ddagger$ Instituto de F\'{\i}sica Te\'orica,
Universidade Estadual Paulista,\\ Rua Pamplona 145, 01405-900 S\~ao Paulo -
SP, Brazil}
\address {$^\S$Institut f\"ur Theoretische Physik, 
J.W. Goethe Universit\"at, Robert Mayer-Stra{\ss}e 10,\\
Postfach 11 19 32, D-60054 Frankfurt/Main, Germany}
\maketitle
\begin{abstract}
We derive the equation of state for hot nuclear matter using Walecka
model in a nonperturbative formalism. We include here the vacuum
polarisation effects arising from the nucleon and scalar mesons 
through a realignment of the vacuum. A ground state structure with
baryon-antibaryon condensates yields the results obtained through
the relativistic Hartree approximation (RHA) of summing baryonic 
tadpole diagrams. Generalization of such a state to include the
quantum effects for the scalar meson fields through the $\sigma$-meson
condensates amounts to summing over a class of multiloop diagrams.
The techniques of thermofield dynamics (TFD) method are used for 
the finite temperature and finite density calculations. The in-medium 
nucleon and sigma meson masses are also calculated in a self consistent 
manner. We examine the liquid-gas phase transition at low temperatures 
($\approx$ 20 MeV), as well as apply the formalism to high temperatures 
to examine for a possible chiral symmetry restoration phase transition.

\end{abstract}

\narrowtext
\def\bfm#1{\mbox{\boldmath $#1$}}

\section{Introduction}
The understanding of hot and dense matter is an interesting and
important problem in the context of heavy ion collision experiments as
well as to study different astrophysical objects such as neutron stars.
Quantum Hadrodynamics (QHD) is a general framework which has been
extensively used to study the nuclear matter, both at zero \cite {walecka}
and finite temperatures \cite{furnst}, as well as to describe the
properties of finite nuclei \cite{finnl}. In the Walecka model (QHD-I), 
the nucleons interact via the scalar ($\sigma$) and vector ($\omega$)
mesons with the scalar and vector couplings fitted from the saturation
density and binding energy of nuclear matter \cite{walecka}.
The hot nuclear matter has been studied neglecting the Dirac sea 
\cite{furnst}, i.e., in the so called no--sea approximation.

To study the nuclear matter at zero temperature including the sea
effects in the relativistic Hartree approximation (RHA), one does
a self consistent sum of the tadpole diagrams for the baryon propagator
\cite {chin}. There have also been calculations including
corrections to the binding energy up to two-loops \cite{twoloop},
which are seen to be rather large as compared to the one-loop
results. However, it is seen that using phenomenological monopole form
factors to account for the composite nature of the nucleons,
such contribution is reduced substantially \cite{prakash}
so that it is smaller than the one-loop result. However, without 
inclusion of such form factors the mean-field theory is not stable 
against a perturbative loop expansion. This might be because
the couplings involved here are too large (of order of 10)
and the theory is not asymptotically free. Hence 
nonperturbative techniques need to be developed to consider
nuclear many-body problems. 

The approximation scheme adopted here is nonperturbative and, 
it uses a squeezed coherent type of 
construction for the ground state \cite{cond,amhm} which amounts 
to an explicit vacuum realignment. The input here is
equal-time quantum algebra for the field operators
with a variational ansatz for the 
vacuum structure and does not use any perturbative expansion
or Feynman diagrams. We have earlier seen that this correctly
yields the results of the Gross-Neveu model \cite{gn} as obtained 
by summing an infinite series of one-loop diagrams. It was also
seen to reproduce the gap equation in an effective QCD
Hamiltonian \cite{chirl} as obtained through the solution of the
Schwinger-Dyson equations for the effective quark propagator.
In an earlier work \cite{mishra}, such a nonperturbative method 
was applied to consider vacuum polarization effects 
in nuclear matter, where it was shown that a realignment of 
the ground state in nuclear matter with baryon--antibaryon condensates
is equivalent to the relativistic Hartree approximation (RHA). We had 
then included the quantum corrections arising from the scalar meson 
in a similar way through sigma condensates, amounting to summimg over
a class of multiloop diagrams. Recently, the formalism has also been 
generalised to include strange baryons and its effects on the 
composition of neutron star matter as well as gross structural 
properties of neutron stars \cite{shm}. In the present
paper, we study hot nuclear matter including the vacuum polarisation
effects arising from the nucleon and scalar meson fields. The method of
thermofield dynamics (TFD) is used here to study the ``ground state"
(the state with minimum thermodynamic potential)
at finite temperature and density. The temperature and density dependent 
baryon and sigma masses are also calculated in a self-consistent
manner in the present framework. We note that in the Walecka model,
TFD has been applied to compute perturbatively quantum corrections to the
temperature dependent Hartree mean field \cite{rhatfd} and effect of 
such corrections on the equation of state. Here we shall however
use a {\em nonperturbative variational} approach to study symmetric
nuclear matter within Walecka model using TFD for small temperatures
(associated with liquid-gas phase transition) as well as at high
temperatures to discuss possible chiral restoration transition.
The ansatz functions involved in such an approach shall be determined
through minimisation of the thermodynamic potential.

We organize the paper as follows. In section II, we study the quantum
correction effects from the nucleon and sigma fields in nuclear matter
as simulated through baryon--antibaryon and scalar meson condensates.
The quantum vacuum in nuclear matter is generalized to finite temperatures
using the thermofield dynamics method. 
The corresponding condensate functions as well as the functions
introduced in the definition of the thermal vacuum are determined 
through the minimisation of the thermodynamic potential. 
This enables us to obtain the equation of state (EOS) and various 
thermodynamic quantities for the hot nuclear matter.
In section III,  we discuss the results obtained in the present
nonperturbative framework of inclusion of quantum corrections.
Unlike the mean field calculations \cite {furnst}, we observe that,
at temperatures of around 200 MeV, the decrease in the nucleon mass is slower
when one includes such quantum correction effects, and need not be
indicative of a chiral symmetry restoration phase transition.
We also study the effect of quantum corrections on the liquid gas 
phase transition. The value of critical temperature for such a 
phase transition is seen to be lowered due to quantum effects
arising from the scalar mesons.  In section IV, we give a brief 
summary of the present work and discuss possible outlook.

\section {Quantum Vacuum in Nuclear Matter}
We shall start with briefly recapitulating the formalism for studying
the nuclear matter including the quantum
correction effects as arising through a realignment of the vacuum
with baryon--antibaryon and scalar meson condensates \cite{mishra}.
The Lagrangian density is given as
\begin{eqnarray}
{\cal L}&=&\bar \psi (i\gamma^\mu \partial_\mu
-M-g_\sigma \sigma-g_\omega\gamma^\mu \omega_\mu)\psi
+\frac{1}{2}\partial^\mu\sigma
\partial_\mu\sigma-\frac{1}{2} m_\sigma ^2 \sigma^2
-\lambda \sigma^4
\nonumber\\
&+&\frac{1}{2} m_\omega^2 \omega^\mu \omega_\mu
-\frac{1}{4}(\partial_\mu \omega_\nu -\partial_\nu \omega_\mu)
(\partial^\mu \omega^\nu -\partial^\nu \omega^\mu).
\end{eqnarray}
In the above, $\psi$, $\sigma$, and $\omega_\mu$ are the fields
for the nucleon, $\sigma$, and $\omega$ mesons
with masses M, $m_\sigma$, and $m_\omega$ respectively. 
The quartic coupling  term in $\sigma$ is necessary for the 
sigma condensates through a vacuum realignment, to exist \cite{mishra}. 
We retain the quantum nature of both the nucleon and the scalar meson 
fields, where as, the vector $\omega$-- meson is treated as a 
classical field, using the mean field approximation for $\omega$--meson,
given as $\langle \omega^\mu \rangle=\delta_{\mu 0} \omega_0$.
The reason is that without any quartic or any other higher order term 
for the $\omega$-meson, the quantum effects generated due to $\omega$-meson 
through the present variational ansatz turns out to be zero.

The Hamiltonian density can then be written as
\begin{equation}
{\cal H}={\cal H}_N+{\cal H}_\sigma+{\cal H}_\omega,
\end{equation}
with
\begin{mathletters}
\begin{equation}
{\cal H}_N=\psi ^\dagger(-i \bf \alpha \cdot \bfm \bigtriangledown
+\beta M)\psi + g_\sigma\sigma \bar \psi\psi,
\end{equation}
\begin{equation}
{\cal H}_\sigma=
\frac {1}{2} {\dot \sigma}^2+
\frac{1}{2} \sigma (-{\bfm \bigtriangledown}^2)
\sigma+\frac{1}{2} m_\sigma ^2 \sigma^2+\lambda \sigma^4,
\label{lwsg}
\end{equation}
\begin{equation}
{\cal H}_\omega= g_\omega \omega_0 \psi
^\dagger \psi
 -\frac{1}{2} m_\omega ^2 \omega_0^2.
\end{equation}
\end{mathletters}
We may now write down the field expansion for the nucleon field
$\psi$ at time $t=0$ as given by \cite{mishra}
\begin{equation}
\psi(\bfm x)=\frac {1}{(2\pi)^{3/2}}\int \left[U_r(\bfm k)c_{Ir}(\bfm k)
+V_s(-\bfm k)\tilde c_{Is}(-\bfm k)\right] e^{i\bfm k\cdot \bfm x} d\bfm k,
\end{equation}
with $c_{Ir}$ and $\tilde c_{Is}$ as the baryon  annihilation 
and antibaryon creation operators with spins $r$ and $s$ respectively,
and $U$ and $V$ are the spinors associated with the particles and
antiparticles respectively \cite{mishra}.
Similarly, we may expand the field operator of the scalar field $\sigma$ 
in terms of the creation and annihilation operators, at time $t=0$ as
\begin{equation}
\sigma (\bfm x,0)={1\over{(2 \pi)^{3/2}} }\int{{d\bfm k\over{\sqrt
{2 \omega (\bfm k)}}}\left(a(\bfm k)+
a^\dagger(-\bfm k)\right)e^{i\bfm k\cdot\bfm x}},
\label{expan}
\end{equation}
In the above, $\omega (\bfm k)=\sqrt{\bfm k^2+m_\sigma^2}$.
The perturbative vacuum is defined corresponding to this basis through 
$a\mid vac\rangle=0=c_{Ir}\mid vac \rangle
=\tilde c_{Ir}^\dagger\mid vac \rangle$. 

To include the vacuum polarisation effects for hot nuclear matter, 
we shall now consider a trial state with baryon--antibaryon and 
scalar meson condensates and then generalize the same to the
finite temperatures and densities \cite{mishra}. 
We thus explicitly take the ansatz for the trial state as 
\begin{equation}
|F\rangle =U_\sigma U_F|vac\rangle,
\label{cond}
\end{equation}
with
\begin{equation}
U_F= \exp \Big[ \int d\bfm k ~f(\bfm k)~{c_{Ir}^\dagger (\bfm k)}~
a_{rs} \tilde c _{Is} (-\bfm k)-h.c. \Big]
\end{equation}
Here $a_{rs}=u_{Ir}^\dagger(\bfm \sigma \cdot \hat k)v_{Is}$ 
and $f(\bfm k)$ is a trial function associated with baryon-antibaryon 
condensates.
For the scalar meson sector,
$ U_\sigma=U_{II}U_{I}$
where $U_{i}=\exp(B_i^\dagger~-~B_i),\,(i=I,II)$. Explicitly the
$B_{i}$ are given as
\begin{equation}
B_I^\dagger=\int {d\bfm k \sqrt{\omega (\bfm k)\over 2}
f_\sigma(\bfm k) a^\dagger(\bfm k)},\;\;\;
{B_{II}}^\dagger={1\over 2}\int d\bfm k g(\bfm k){a'}^\dagger(\bfm k)
{a'}^\dagger(-\bfm k). 
\end{equation}
In the above, $a'(\bfm k)=U_I a(\bfm k) U_I^{-1}=a(\bfm k)-
\sqrt{\frac{\omega (\bfm k)}{2}}f_\sigma(\bfm k)$ corresponds to a
shifted field operator associated with the coherent state 
\cite{mishra,amhm} 
and satisfies the usual quantum algebra. Further, to preserve 
translational invariance $f_\sigma(\bfm k)$ has to be  proportional to $\delta
(\bfm k)$ and  we take $f_\sigma(\bfm k)=\sigma _{0} (2\pi)^{3/2}
\delta (\bfm k)$. $\sigma_0$ corresponds to a
classical field of the conventional approach \cite{mishra}.
Clearly, the ansatz state is not annihilated by the operators,
$c$, $\tilde c^\dagger$ and $a$. However, one can define operators,
$d$, $\tilde d^\dagger$ and $b$, related through a Bogoliubov
transformation to these operators, which will annihilate 
the state $|F\rangle$.


We next use the method of thermofield dynamics \cite{tfd}
to construct the ground state for nuclear matter
at finite temperature. Here the statistical average of an operator 
is written as an expectation value with respect to a `thermal vacuum' 
constructed from operators defined on an extended Hilbert space. 
The `thermal vacuum' is obtained from the zero 
temperature ground state through a thermal Bogoliubov transformation.
We thus generalise the state, as given by (\ref{cond}) to finite 
temperature and density as
\cite{mishra,amhm}
\begin{equation}
|F,\beta\rangle=U_\sigma(\beta)U_F(\beta)|F\rangle.
\label{trialansatz}
\end{equation}
The temperature-dependent unitary operators $U_\sigma(\beta)$ and
$U_F(\beta)$ are given as \cite{tfd}

\begin{equation}
U_\sigma(\beta)=\exp{\Bigg({1\over 2}\int d\bfm k \theta_\sigma(\bfm k,\beta) 
b^\dagger(\bfm k) {\underline b}^\dagger(-\bfm k)-h.c.\Bigg)}. 
\end{equation}
and 
\begin{equation}
U_F(\beta) =\exp{\Bigg(
\int d\bfm k ~\bigg[\theta_-(\bfm k,\beta)~
d_{Ir}^\dagger (\bfm k)~{\underline d}_{Ir}^\dagger(-\bfm k)
+\theta_+(\bfm k,\beta)~ \tilde d_{Ir}(\bfm k)~\tilde 
{\underline d}_{Ir}(-\bfm k)\bigg]-h.c.\Bigg)}.
\end{equation}
The underlined operators are the operators corresponding to the
doubling of the Hilbert space that arise in thermofield dynamics method.
We shall determine the condensate functions $f(\bfm k)$ and $g(\bfm k)$,
and the functions  $\theta_\sigma(\bfm k,\beta)$, $\theta_-(\bfm k,\beta)$ 
and $\theta_+(\bfm k,\beta)$ 
of the thermal vacuum through minimisation of the thermodynamic potential.
The thermodynamic potential is given as
\begin{equation}
\Omega \equiv -p=\epsilon-\frac{1}{\beta}{\cal S}-\mu \rho_B,
\label{thermpot}
\end{equation}
where $\epsilon$ and ${\cal S}$ are the energy- and entropy-
densities of the thermal vacuum, and $\rho_B$ is the baryon density.
On evaluation, the energy density of the thermal vacuum is given as,
\begin{equation}
\epsilon\equiv\langle {\cal H}\rangle_\beta =\epsilon_N
+\epsilon_\omega+\epsilon_\sigma
\end{equation}
with 
\begin{mathletters}
\begin{equation}
\epsilon_N=-\frac{\gamma}{(2\pi)^3}\int d \bfm k \Bigg[
\epsilon (\bfm k)\cos 2f(\bfm k)+\frac{g_\sigma \sigma_0}{\epsilon (k)}
\bigg(M \cos 2f(\bfm k)+|\bfm k|
\sin2 f (\bfm k)\bigg)\Bigg] (\cos^2 \theta_+ -\sin^2 \theta _-),
\end{equation}
\begin{equation}
\epsilon_\omega=g_\omega \omega_0 ~ \gamma (2\pi)^{- 3}\int d \bfm k 
(\cos^2 \theta_+ +\sin^2 \theta _-)-\frac{1}{2}m_\omega^2 \omega_0^2,
\end{equation}
and 
\begin{eqnarray}
\epsilon_\sigma &=&
\frac{1}{2}{1 \over {(2 \pi)^3}} \int{d \bfm k
\over 2\omega (k)}\Bigg[ k^{2}(\sinh\!2g +\cosh\!2g)
+\omega^2 (k)(\cosh\!2g -\sinh\!2g)\Bigg]\cosh\!2\theta_\sigma(\bfm k,\beta)
\nonumber\\
&+& \frac{1}{2}m_\sigma^2I(\beta)+6\lambda\sigma_0^2 I(\beta)+
3\lambda I(\beta)^2
+\frac{1}{2}m_\sigma^2\sigma_0^2 +\lambda \sigma_0^4,
\label{en}
\end{eqnarray}
\end{mathletters}
with
\begin{equation}
I(\beta)={1 \over (2 \pi)^3}\int{{d\bfm k \over {2\;\omega (k)}}
(\cosh \!2g +\sinh\!2g)}\cosh\!2\theta_\sigma(\bfm k,\beta).
\label{gep9c}
\end{equation}
The entropy density
\begin{eqnarray}
{\cal S}&=&-\gamma (2\pi)^{-3} \int d \bfm k \bigl[
\sin^2 \! \theta_- \ln(\sin^2\!\theta_-)+
\cos^2 \! \theta_- \ln(\cos^2\!\theta_-)\nonumber\\
&+&
\sin^2 \! \theta_+ \ln(\sin^2\!\theta_+)+
\cos^2 \! \theta_+ \ln(\cos^2\!\theta_+)
\bigr]\nonumber\\
&+&
(2\pi)^{-3} \int d \bfm k \bigl[
\cosh^2 \! \theta_\sigma \ln(\cosh^2\!\theta_\sigma)
- \sinh^2 \! \theta_\sigma \ln(\sinh^2\!\theta_\sigma)
\bigr]+{\cal S}_\omega
\label{entr}
\end{eqnarray}
and the baryon density
\begin{equation}
\rho_B=\gamma (2\pi)^{- 3}\int d \bfm k 
(\cos^2 \theta_+ +\sin^2 \theta _-).
\end{equation}
In the above, $\gamma$  is the spin isospin degeneracy factor 
and is equal to $4$ for nuclear matter. Further, ${\cal S}_\omega$ is
the contribution to the entropy density from
$\omega$-meson.
Extremising the thermodynamic potential, $\Omega$ with respect
to the condensate function $f(\bfm k)$ and the functions
$\theta_{\mp}$ corresponding to the nucleon sector yields
\begin{equation}
\tan 2 f(\bfm k)=\frac{g_\sigma \sigma_0 |\bfm k|}{\epsilon(k)^2+
M g_\sigma \sigma_0}
\end{equation}
and
\begin{equation}
\sin^2 \theta _{\mp}=\frac{1}{\exp(\beta(\epsilon^*(k)\mp \mu^{*})) 
+1},
\label{distr}
\end{equation}
with $\epsilon^*(k)=(k^2+{M^*}^2)^{1/2}$ and $\mu^{*}=\mu -g_\omega
\omega_0$ as the effective energy and effective chemical
potential, where the effective nucleon mass $M^{*}=M+g_\sigma \sigma_0$.

For the sigma meson sector, on extremising the thermodynamic potential,
the functions $g(\bfm k)$ and $\theta_\sigma$ are obtained as

\begin{equation}
\tanh\!{2 g(k)}=-\,{{6 \lambda I(\beta)+6 \lambda {\sigma _0}^2}\over {
{\omega (k)}^{2}+6 \lambda I(\beta)+6 \lambda {\sigma _{0}}^{2}}}
\label{gk}
\end{equation}
and
\begin{equation}
\sinh^2 \theta _\sigma=\frac{1}{e^{\beta\omega'(\bfm k,\beta) -1}};
\;\;\;
\omega'(\bfm k,\beta)=(\bfm k^2+M_\sigma(\beta)^2)^{1/2}.
\end{equation}
In the above, the effective scalar meson mass is given as
\begin{equation}
M_\sigma(\beta)^2=m_\sigma^2+12\lambda I(\beta) +12\lambda \sigma_0^2
\label{m2}
\end{equation}
with
\begin{equation}
I(\beta)=\frac{1}{(2\pi)^3}\int\frac{d\bfm k}{2} 
\frac{1}{(\bfm k^2+M_\sigma(\beta)^2)^{1/2}}
\label{I}
\end{equation}
It is clear from the equation (\ref{gk}) that in the absence of a 
quartic coupling no such condensates are favoured since the
condensate function, $g(\bfm k)$ vanishes for $\lambda=0$.

Then the expression for the energy density becomes
\begin{equation}
\epsilon=\epsilon_N+\epsilon_\omega+\epsilon_\sigma,
\end{equation}
with
\begin{mathletters}
\begin{equation}
\epsilon_N=\gamma (2\pi)^{-3}\int d \bfm k (k^2+{M^*}^2)^{1/2} 
(\sin^2 \theta_- -\cos ^2 \theta_+) ,
\end{equation}
\begin{equation}
\epsilon_\omega=
 g_\omega \omega_0 ~\gamma (2\pi)^{-3}\int d \bfm k 
(\sin^2 \theta_-+\cos^2 \theta_+)
-\frac{1}{2}m_\omega^2 \omega_0^2,
\end{equation}
\begin{equation}
\epsilon_\sigma=
{1 \over {2}}m_\sigma^2{\sigma _0}^2\,+\lambda {\sigma _0}^4
+\frac{1}{2}\frac{1}{(2\pi)^3}\int d\bfm k (k^2+M_\sigma(\beta)^2)^{1/2}\cosh 
2\theta_\sigma -3 \lambda I(\beta)^2.
\label{pot}
\end{equation}
\end{mathletters}

After subtracting out the vacuum contributions ($\theta_{\pm}$=0,
$f$=0 part) for the nucleon sector, one obtains,
\begin{equation}
\Delta \epsilon_N=\epsilon_{finite}^{N}
+\Delta \epsilon,
\end{equation}
where,
\begin{equation}
\epsilon_{finite}^{N}=
\gamma (2\pi)^{-3}\int d \bfm k (k^2+{M^*}^2)^{1/2} 
(\sin^2 \theta_- +\sin ^2 \theta_+) 
\end{equation}
and
\begin{equation}
\Delta \epsilon =-\gamma (2\pi)^{-3}\int d\bfm k 
\Big [(\bfm k^2 +{M^*}^2)^{1/2} -(\bfm k^2 +M^2)^{1/2} \Big],
\label{chindiv}
\end{equation}

The contribution arising from the Dirac sea effect given by (\ref{chindiv}),
is identical to that of summing over the baryonic tadpole diagrams
of RHA, before renormalisation. This is renormalised by adding the 
counter terms, as for the zero temperature situation, given as
\begin{equation}
\epsilon_{ct}= \sum_{n=1}^4 C_n \sigma_0^n,
\end{equation}
since finite temperature does not introduce any fresh divergences.

For the $\omega$-sector, the pure vacuum contribution to $\rho_B$
is subtracted out, which amounts to a normal ordering of the number
operator $\psi ^\dagger \psi$. This yields the usual energy density
for the $\omega$-sector, 

\begin{equation}
\epsilon_{\omega}=
g_\omega \omega_0 \rho_B^{ren}-\frac {1}{2} m_\omega^2 \omega_0^2,
\end{equation}
with
\begin{equation}
\rho_B^{ren}=\gamma (2\pi)^{-3}\int d\bfm k 
(\sin^2 \theta_- -\sin ^2 \theta_+).
\end{equation}

For the $\sigma$ sector, since there are no additional divergences
arising from finite temperatures, we adopt the same renormalization 
procedure as in Ref. \cite{mishra}. This yields
the gap equation for  field dependent effective sigma mass, 
$M_\sigma(\beta)$, in terms of the renormalised parameters as
\begin{equation}
M_\sigma(\beta)^2=m_R^2+12\lambda_R\sigma_0^2+12\lambda_R I_f(M_\sigma(\beta)),
\label{mm2}
\end{equation}
where
\begin{equation}
I_f(M_\sigma(\beta))=\frac{M_\sigma(\beta)^2}{16\pi^2}
\ln \Big(\frac{M_\sigma(\beta)^2}{m_R^2} \Big)+
\frac{1}{(2\pi)^3}\int d\bfm k \frac{\sinh^2\theta_\sigma(\bfm k,
\beta)}{(\bfm k^2+ M_\sigma(\beta)^2)^{1/2}},
\label{if}
\end{equation}
Simplifying equation (\ref{pot}) and subtracting the vacuum contribution, 
we obtain the energy density for the $\sigma$,
\begin{eqnarray}
\Delta \epsilon_\sigma 
&=& \frac{1}{2} m_R^2 \sigma_0^2+ 3\lambda_R \sigma_0^4 
+\frac {M_\sigma^4}{64\pi^2}
\Biggl(\ln\Big(\frac{M_\sigma^2}{m_R^2}\Big)-\frac{1}{2} \Biggr)
-3\lambda_R I_f^2\nonumber\\
&-&\frac {M^4_{\sigma,0}}{64\pi^2}
\Biggl(\ln\Big(\frac{M_{\sigma,0}^2}{m_R^2}\Big)-\frac{1}{2} \Biggr)
+3\lambda_R I_{f0}^2,
\label{vph0}
\end{eqnarray}
where $M_{\sigma,0}$ and $I_{f0}$ are the expressions
as given by eqs. (\ref{mm2}) and (\ref{if}) with $\sigma_0=0$.
We might note here that the gap equation given by (\ref{mm2})
is identical to that obtained through resumming the daisy 
and superdaisy graphs \cite{pi} and hence includes
higher order corrections from the scalar meson field.

In the absence of the quartic interaction ($\lambda_R=0$), equation
(\ref{vph0}) reduces to
\begin{equation}
\Delta \epsilon_\sigma 
= \frac{1}{2} m_R^2 \sigma_0^2,
\end{equation}
which refers to the RHA. Also, we note that the sign of $\lambda_R$ 
must be chosen to be positive, because otherwise the energy density
would become unbounded from below with vacuum fluctuations 
\cite{serot2,fox,furnstahl}.

One then obtains the renormalised energy density as
\begin{equation}
\epsilon_{ren}=\epsilon_{finite}^{(N)}
+\Delta \epsilon_{ren}+\epsilon_\omega+\Delta \epsilon_\sigma,
\end{equation}
with,
\begin{eqnarray}
\Delta \epsilon_{ren} &=& -\frac{\gamma}{16\pi^2}
 ( {M^*}^4 \ln \Big (\frac{M^*}{M}\Big )
+M^3 (M-M^*)-\frac{7}{2} M^2 (M-M^*)^2 \nonumber\\
& + & \frac{13}{3} M (M-M^*)^3 -
 \frac{25}{12} (M-M^*)^4 )
\end{eqnarray}
\noindent as the contribution from the Dirac sea.
The thermodynamic potential, $\Omega$, given  by equation 
(\ref {thermpot}), after subtracting out the vacuum contributions,
is now finite and is given in terms of the meson fields,
$\sigma_0$ and $\omega_0$.

Extremisation  of the thermodynamic potential with respect
to the meson fields $\sigma_0$ and $\omega_0$ give the 
self--consistency conditions for 
$\sigma_0$ (and hence for the effective nucleon mass,
$M^*=M+g_\sigma \sigma_0$), as
\begin{mathletters}
\begin{equation}
\frac {d(\Delta \epsilon_\sigma)}{d\sigma_0}
+\frac {\gamma}{(2\pi)^3}g_\sigma \int d \bfm k 
\frac {M^*}{(\bfm k^2+{M^*}^2)^{1/2}}
(\sin^2 \theta_- +\sin ^2 \theta_+)
+\frac {d(\Delta \epsilon_{ren})}{d\sigma_0}=0
\label{selfsig}
\end{equation}
and, for the vector meson field, $\omega_0$, as
\begin{equation}
\omega_0=\frac {g_\omega}{m_\omega^2}\frac {\gamma}{(2\pi)^3}
\int d \bfm k (\sin^2 \theta_- -\sin ^2 \theta_+)
\label{selfomg}
\end{equation}
\label{selfcons}
\end{mathletters}
where $\sin^2 \theta_{\mp}$ as the thermal distribution functions
for the baryons and antibaryons, given through equation (\ref{distr}).

\section {Results and Discussions}

We now proceed with the finite temperature calculations for the
nuclear matter. The values for $C_s^2=g_{\sigma}^2 M^2/m_\sigma^2$ 
and $C_v^2=g_{\omega}^2 M^2/m_\omega^2$, are given as
$C_s^2$=183.3,167.5,137.9 and $C_v^2$=114.7,96.45,63.7, 
for the RHA and with quantum corrections from
sigma meson, for $\lambda_R$=1.8 and 5 respectively \cite{mishra}.
These values were fitted from the nuclear matter saturation properties as 
$\rho_0=0.193 fm^{-3}$ and binding energy as $-$15.75 MeV.
For given values of the temperature and the baryon chemical potential,
$\mu$, we calculate the different thermodynamic quantities,
with the meson fields determined self consistently
from equation (\ref{selfcons}).

We plot the  pressure as a function of the baryon density, $\rho_B$ 
for low temperatures to examine the liquid gas phase transition,
in figures 1 and 2, for the RHA and $\lambda_R$=1.8 respectively,
At zero temperature, the pressure decreases with density, reaches
a minimum, then increases and passes through $p=0$ at $\rho=\rho_0$,
where the binding energy per nucleon is a minimum. The negative
pressure indicates a mechanical instability of the uniform nuclear 
matter of saturation density, $\rho_0$. This has however an 
interesting physical interpretation. For the densities below the 
saturation density, the uniform nuclear matter is unstable and will 
break up into regions of nuclear matter with density, $\rho=\rho_0$ 
and zero pressure, surrounded by regions of vacuum with $\rho=0$ 
and $p=0$ \cite{buballa}. This pocket in the pressure versus density 
curve, disappears at the liquid gas phase transition point. For RHA, 
this transition appears to occur at around 23 MeV. Inclusion of sigma 
condensates reduces this critical temperature, $T_c$ to 22 MeV for 
$\lambda_R$=1.8 and to 21 MeV for $\lambda_R$=5.

For higher temperatures (beyond the liquid gas phase transition),
we plot the EOS for RHA and for $\lambda_R$=1.8 in figures 3 and 4
respectively. These plots show a softening of the
EOS due to quantum effects arising from the scalar meson sector.
Such a softening of the EOS with  inclusion of the quantum
corrections was already observed for the zero temperature
situation \cite{mishra}, which had given rise to a lower value
for the incompressibility. For a given $\rho_B$, the pressure
has the usual trend of  increasing with temperature \cite{furnst}.
The pressure for $\rho_B=0$ is seen to be already nonzero and
appreciable at around a temperature of 150 MeV for RHA and 
200 MeV for $\lambda_R$=1.8. This has contributions arising 
from the thermal distributions of baryons and antibaryons, 
as well as from a nonzero value for the sigma field.

The magnitude of the scalar meson field, $\sigma$ as obtained 
through the self-consistency condition (\ref{selfsig}), is plotted
as a function of the baryon density for various temperatures 
in figure 5 for $\lambda_R$=1.8.
It may be noted that for $\rho_B$=0, $\sigma$ field becomes 
nonzero at a temperature of around 160 MeV due to thermal effects, 
and is observed to have attained an appreciable value at 200 MeV 
due to contributions from the thermal distribution functions. 
The sigma field attaining a nonzero value was also observed
for nuclear matter in the mean field approximation in
 Walecka model \cite{furnst}, which had led to a sharp fall 
in the effective nucleon mass between 150 MeV and 200 MeV. 
The rapid fall of $M_N^*$ with increasing temperature, was
ascribed to resemble the situation, when the system becomes 
a dilute gas of baryons in a sea of baryon-antibaryon pairs. 

The density and temperature dependent nucleon mass is plotted 
in figures 6 and 7, for RHA and $\lambda_R$=1.8 respectively. 
The quantum effects are seen to increase the effective nucleon mass. 
It may be noted that the in-medium baryon mass increases with 
larger value of the quartic coupling. Hence, with inclusion of 
quantum effects, the rapid fall in effective nucleon mass as 
observed in the mean field calculations \cite {furnst} is not 
seen in the present calculation. 
The nucleon mass here becomes about half of its vacuum
value at T=250 MeV, as compared to $M^*/M \approx 0.2$ in Ref.
\cite{furnst} at T=200 MeV.

The dependence of the entropy density on the temperature and
density for $\lambda_R$=1.8 is shown in figure 8. This becomes
nonzero for zero baryon density at around a temperature
of 160 MeV, with contributions from the nonzero value for the
sigma field.  However, the increase with temperature, for
$\rho_B$=0, is rather gradual and need not be associated with
a phase transition. A similar behaviour of entropy density
becoming nonzero at higher temperatures,  for $\rho_B$=0, was 
also observed in the Walecka model mean field calculations 
\cite {furnst} as well as in the QMC model \cite{qmct}.

Finally, the inclusion of quantum corrections through scalar meson
condensates enables us to determine the in-medium sigma meson
mass in a self--consistent manner, which we plot in figure 9 for
$\lambda_R$=1.8. An increase in the quartic coupling, $\lambda_R$ 
increases the value for the effective sigma mass \cite{mishra}.

\section {Summary}

To summarize, in the present work, we have studied the hot nuclear matter
taking into account the vacuum polarisation effects within a nonperturbative
variational calculation. The approximation here lies in the ansatz for 
the ground state. A realignment of the ground state with baryon-antibaryon 
condensates takes into account the summing over baryonic tadpole diagrams. 
Generalization of the vacuum with mesons gives rise to summing over 
multi-loop diagrams. We could have generalized the ansatz to include vector 
meson ($\omega$)-condensates in a similar way. However, such an
ansatz with the present Lagrangian would lead to a trivial solution
for the condensate function (thus giving us the mean field result). 
However, with a quartic term \cite{bodmer} for the $\omega$ meson, 
the present ansatz for the ground state would give rise to a nontrivial
solution for the corresponding gap equation. Including such a term, 
it would be interesting to study the quantum effects arising from 
the vector meson. 

We would also like to mention that the thermal distribution and the condensate
functions are obtained here through a {\em minimisation} of the thermodynamic
potential which included interactions with the corresponding effective
masses for the $\sigma$-meson and nucleon determined self--consistently
through the equations (\ref{mm2}) and (\ref{selfsig}) at a given temperature.

We have looked into the liquid-gas phase transition, and observed that
the critical temperature for the phase transition is lowered due to quantum
correction effects from the scalar meson sector. At high temperatures, 
with quantum effects, the effective nucleon mass does not decrease 
as rapidly as in the mean field results \cite{furnst}. With increase in
temperature, the decrease of nucleon mass as well as the increase in
the entropy density appear to be rather gradual, i.e., without any sudden
change and is not indicative of any (chiral or otherwise) phase transition. 

\section{Acknowledgement}
One of the authors (P. K. P.) would like to acknowledge 
FAPESP (Processo-99/08544-0) for financial support and
the IFT, S\~ao Paulo, for kind hospitality. This work was 
initiated when two of the authors (A. M. and P. K. P.) were visiting
Institut f\"ur Theoretische Physik, J.W. Goethe Universit\"at, 
Frankfurt, and they would like to thank the Institut 
f\"ur Theoretische Physik, for providing facilities 
and to Alexander von Humboldt foundation for financial support
during that period. AM would like to acknowledge many useful
discussions with Prof. J. C. Parikh.

\begin{figure}
\psfig{file=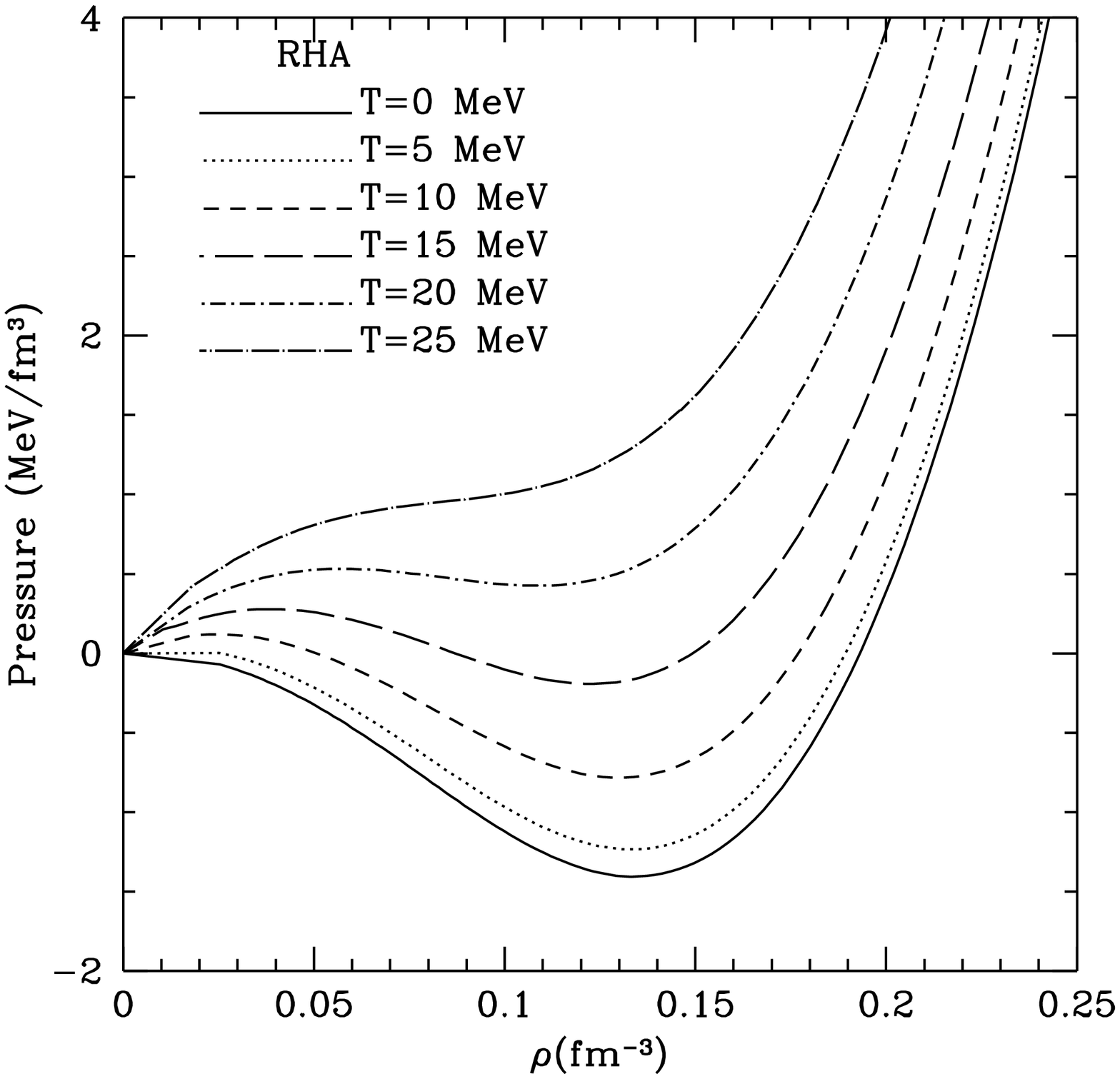,width=9cm,height=9cm}
\caption{Pressure versus the baryon density for RHA.
The disappearance of the pocket at temperature of around 23 MeV 
is indicative of a liquid-gas phase transition.} 
\end{figure}
\begin{figure}
\psfig{file=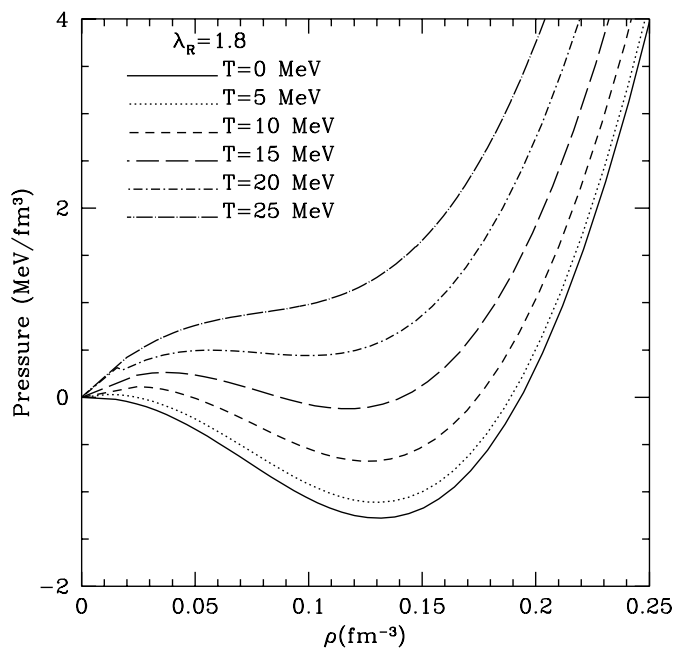,width=9cm,height=9cm}
\caption{Pressure versus the baryon density for $\lambda_R=1.8$.
Quantum corrections due to scalar meson sector gives rise to a 
smaller value of $T_c$ of 22 MeV as compared to that of RHA.}
\end{figure}
\begin{figure}
\begin{center}
\psfig{file=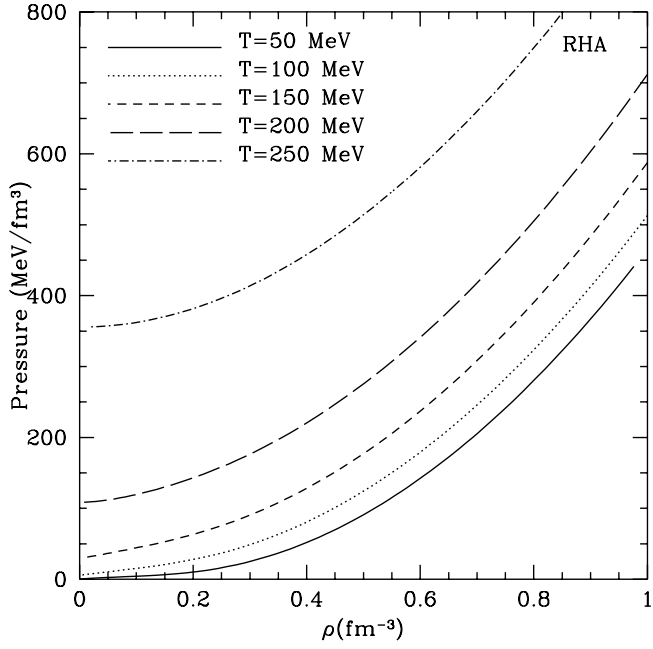,width=9cm,height=9cm}
\end{center}
\caption{The equation of state for hot nuclear matter for the relativistic
Hartree approximation.}
\end{figure}
\begin{figure}
\begin{center}
\psfig{file=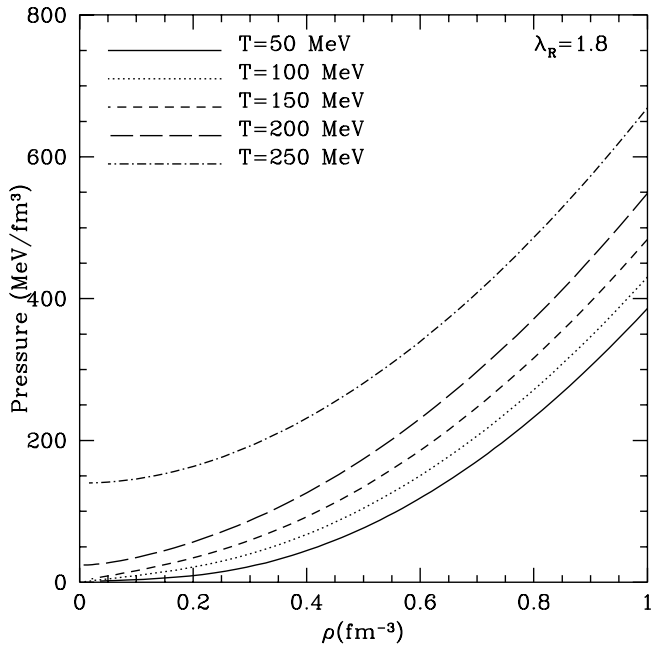,width=9cm,height=9cm}
\end{center}
\caption{The equation of state for $\lambda_R=1.8$.
The quantum correction from scalar mesons gives a softer equation 
of state.}
\end{figure}
\begin{figure}
\psfig{file=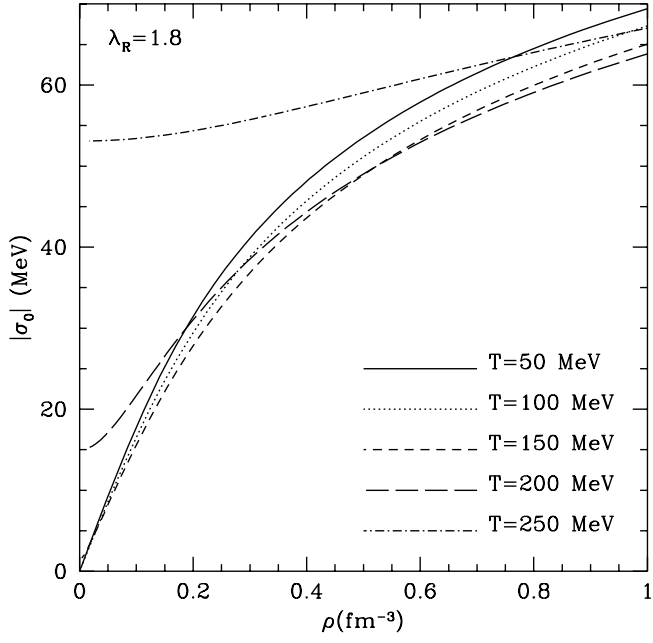,width=9cm,height=9cm}
\caption {The meson field strengths as functions of baryon density.}
\end{figure}
\begin{figure}
\begin{center}
\psfig{file=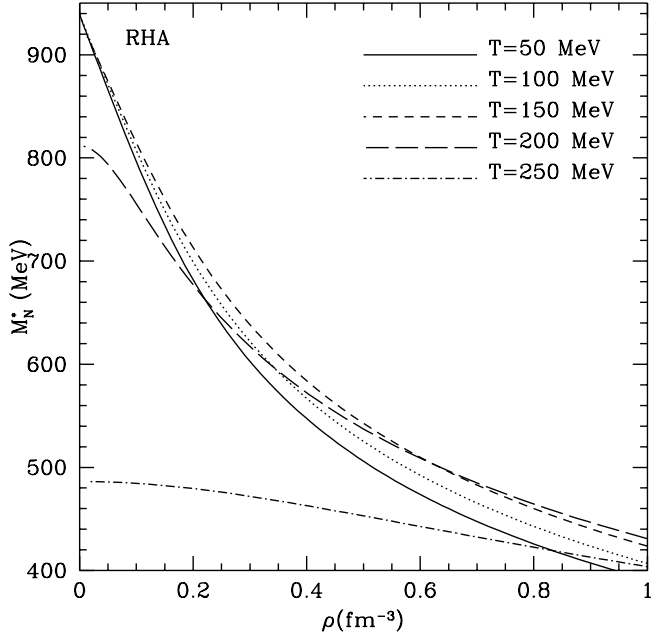,width=9cm,height=9cm}
\end{center}
\caption{Effective baryon masses in the medium for RHA.}
\end{figure}
\begin{figure}
\begin{center}
\psfig{file=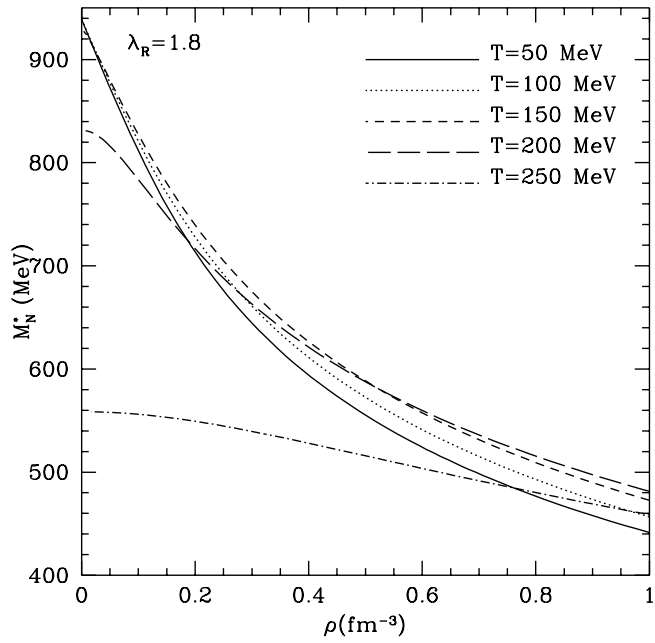,width=9cm,height=9cm}
\end{center}
\caption{Effective baryon masses in the medium for $\lambda_R=1.8$.
The in-medium baryon mass increases with quantum correction effects.} 
\end{figure}
\begin{figure}
\psfig{file=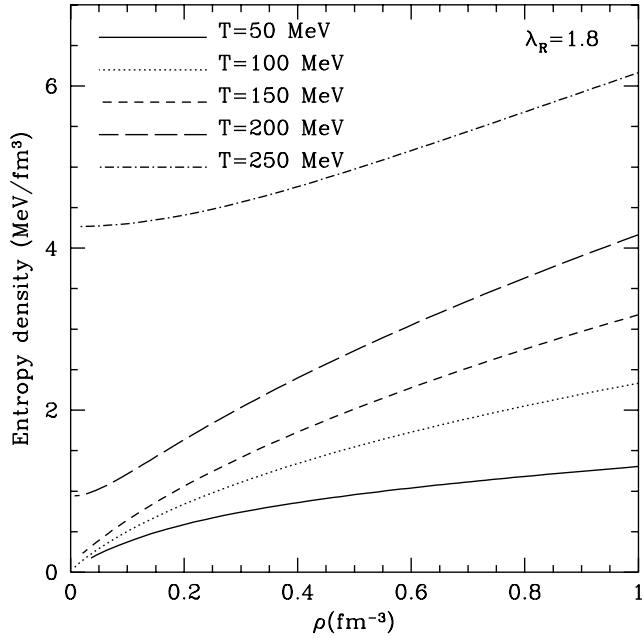,width=9cm,height=9cm}
\caption{Entropy density versus baryon density
for $\lambda_R=1.8$.}
\end{figure}
\begin{figure}
\psfig{file=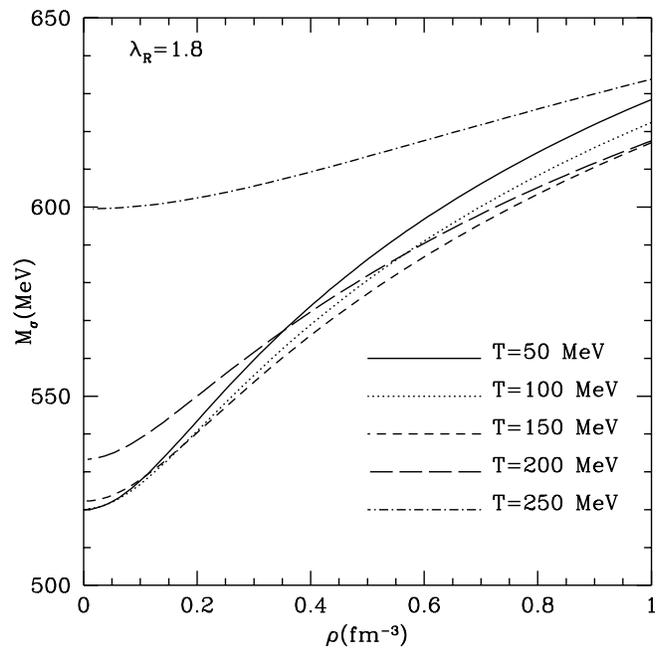,width=9cm,height=9cm}
\caption{In medium scalar meson mass versus baryon density
for $\lambda_R=1.8$.}
\end{figure}
\end{document}